\documentclass[12pt]{article}
\pdfoutput=1
\usepackage{textcomp}
\usepackage{color}
\usepackage{putex}
\usepackage{autobreak}
\usepackage{indentfirst}
\usepackage{graphicx}
\usepackage{float}
\graphicspath{{plots/}}
\usepackage{tabularx}
\usepackage{caption}
\usepackage{amsmath}
\numberwithin{equation}{section}
\usepackage{cancel}
\usepackage{array}
\usepackage{subcaption}
\usepackage{epstopdf}
\usepackage{enumerate}
\usepackage{cite}
\usepackage{youngtab}
\usepackage{tensor}
\usepackage{slashed}
\usepackage[aligntableaux=center]{ytableau}
\usepackage{CJKutf8}
\usepackage[utf8]{inputenc}
\usepackage{rotating}
\usepackage{multirow}
\usepackage[
      colorlinks=true,
      linkcolor=red,
      urlcolor=red,
      filecolor=black,
      citecolor=red,
      ]{hyperref}
\usepackage[noabbrev]{cleveref}
\begin{document}
\institution{forbiddencity}{Kavli Institute for Theoretical Sciences, University of Chinese Academy of Sciences, \cr Beijing 100190, China.}
\title{
On the positivity of the hypergeometric Veneziano amplitude
}
\authors{
Konstantinos C. Rigatos\worksat{\forbiddencity},\footnote{{\hypersetup{urlcolor=black}\href{mailto:rkc@ucas.ac.cn}{rkc@ucas.ac.cn}}}
}
\abstract{
Recently, an infinite family of one-parameter generalisations of the Veneziano amplitude were bootstrapped using as input assumptions an integer mass spectrum, crossing symmetry, high-energy boundedness, and exchange of finite spins. This new result was dubbed the hypergeometric Veneziano amplitude, with a real-valued deformation parameter r. For concreteness we work in a setup where the lowest-mass state is a tachyon of mass $m^2_0=-1$ and using the partial-wave decomposition and the positivity of said decomposition's coefficients we are able to bound the deformation parameter to $r \geq 0$ and, also, to obtain an upper bound on the number of spacetime dimensions $D \leq 26$, which is the critical dimension of bosonic string theory.  
}
\date{\today}
\maketitle
{
\hypersetup{linkcolor=black}
\tableofcontents
}
\newpage
\section{Prologue}\label{sec: prologue}
Bootstrapping scattering amplitudes of massless and massive particles, see \cite{Kruczenski:2022lot} for a recent summarized exposition to advances and developments of the S-matrix bootstrap, is an old theme in theoretical high-energy physics. The idea behind it is to provide an alternative approach to understanding and examining physics theories. Concretely, we can formulate specific mathematical questions about the S-matrix and attempt to answer this kind of questions, rather than resorting to Lagrangian descriptions and sophisticated geometrical approaches. This, in turn, implies that we can understand the theories of interest as being fixed by constraints and conditions that are imposed on scattering amplitudes.

In order to employ any bootstrap algorithm, we have to choose a set of assumptions and conditions and then impose them on a landscape of objects. For the purposes of bootstrapping scattering amplitudes such a set can consist of \textit{crossing symmetry, polynomial residues, and high-energy boundedness}.

An explicit four-point amplitude that satisfies the above and is a meromorphic function, except for its simple poles, was constructed by Veneziano \cite{Veneziano:1968yb} and is given by:
\begin{equation}
\mathcal{M}(s,t) = \frac{\Gamma(-(1+\alpha^{\prime}s))\Gamma(-(1+\alpha^{\prime}t))}{\Gamma(-(1+\alpha^{\prime}s)-(1+\alpha^{\prime}t))}
\,				.
\end{equation}
Today we know, of course, that it describes the $2 \rightarrow 2$ scattering of open-string tachyons of mass $\alpha^{\prime}m^2_0=-1$\footnote{we work with conventions in which $\alpha^{\prime}=1$ for open string theory.}. 

It is worthwhile stressing that a scattering amplitude violating the requirement of tame ultraviolet behaviour is an indication for the breakdown of unitarity and causality of the theory \cite{Camanho:2014apa,Haring:2022cyf}. Very robust expressions have been derived describing bounds for theories that are gapped; the Regge and the Froissart bounds. 

The Veneziano amplitude has been studied quite extensively, with the recent works of \cite{Maity:2021obe,Arkani-Hamed:2022gsa} focusing on the coefficients of the partial-wave decomposition of the amplitude and discussing its unitarity from tree-level considerations, as well as the critical dimension of string theory. Questions regarding its uniqueness were posed since the early days of its discovery. 

Along this particular line of investigation, an answer was given in \cite{Coon:1969yw,Baker:1970vxk,Coon:1972qz} that is nowadays known as the Coon amplitude. This is another amplitude that satisfies the criteria of polynomial boundedness, finite-spin exchange, and meromorphicity. It comes with logarithmic Regge trajectories, and for many years it was disregarded by virtually everyone. Recently, however, it has received revived activity \cite{Jepsen:2023sia,Bhardwaj:2022lbz,Li:2023hce,Figueroa:2022onw,Chakravarty:2022vrp,Geiser:2022icl,Bhardwaj:2023eus}. The Coon amplitude is a deformation of the Veneziano in terms of one-parameter and it exhibits a mass spectrum with discrete levels converging to an infinite density at an accumulation point that is followed by a branch cut.

While the works of \cite{Jepsen:2023sia,Caron-Huot:2016icg} raise concerns regarding the status of unitarity of the Coon amplitude, it was realised in \cite{Maldacena:2022ckr} that string theory admits amplitudes behaving like that since they arise from the of open-string scattering with open strings having their endpoints on a D-brane in AdS.

With the Coon amplitude being able to provide us with an explicit and consistent generalisation of the Veneziano amplitude, people have been revisiting the question of constructing more general four-point amplitudes that are consistent with the principles of the S-matrix bootstrap \cite{Geiser:2022exp,Cheung:2022mkw,Cheung:2023uwn,Cheung:2023adk}. This can, also, be phrased as a question to the uniqueness of string theory. Phrased in simple terms, since string theory amplitudes satisfy particular constraints, are they the only objects doing so?  

In this work we focus our attention on an infinite generalisation of the Veneziano amplitude, recently derived in \cite{Cheung:2023adk}. Similarly to the Coon amplitude, it is, also, written as a one-parameter deformation and has been dubbed the \textit{hypergeometric Veneziano amplitude}; this name will become perfectly clear in the next section. Using the partial-wave decomposition of the amplitude, we wish to impose the positivity of the coefficients in order derive bounds on the allowed values for the parameter $r$ and the spacetime dimensions $D$\footnote{The authors of \cite{Cheung:2023adk} have discussed constraints and bounds resulting from unitarity using a numerical evaluation of the partial-wave coefficients. More specifically, starting from the integral representation of the coefficients, they were able to re-express them as a double-sum and upon an explicit numerical evaluation they were able to constrain the allowed values for the deformation parameter, $r$, and the number of allowed spacetime dimensions, $D$. Our analysis is a more systematic examination of the partial-wave coefficients.}. 

The approach we take here, in order to derive bounds on the allowed values of the deformation parameter $r$ and the spacetime dimension $D$, is to examine the positivity of the coefficients of the partial-wave decomposition of the amplitude. To do so, we compute the residues of the amplitude at the location of its poles. Then, we proceed to decompose the residues in a basis spanned by Gegenbauer polynomials, which is valid for any number of spacetime dimensions, $D \geq 3$\footnote{In the special cases of $D=4$ and $D=5$ the Gegenbauer polynomials reduce to the Legendre polynomials and the Chebyshev polynomials of the second kind.}. As we have already mentioned, the task at hand is to find the numbers that multiply these polynomials, since their non-negativity is tied to the unitarity of the underlying theory. We proceed by utilizing the orthogonality relations that these polynomials obey, and we obtain a relation for the partial-wave coefficients as an integral of those special polynomials and some non-trivial function. Using the generating functions for the special polynomials we can define a ``pseudo-generating function''. After some algebraic manipulations, which consist of writing our expressions as power series expansions we, effectively, have two polynomial expansions for the original equation of the partial-wave coefficients. From that we can read off the terms of appropriate scaling in order to derive the coefficients.     

In addition to the above, we also resort to some experimental guess-work, in order to derive additional analytic expressions for the partial-wave coefficients of sub-leading Regge trajectories. This means that, starting from the original expression for the partial-wave coefficients given as an integral of the special polynomials times some non-trivial function, we compute the integral for some values, we make a guess for the general form of the answer and we proceed to verify our claim by checking explicitly some non-trivial values\footnote{In this context, by non-trivial we mean some values that were not used in order to claim the answer of the partial-wave coefficients.}.

The main results of this work is the derivation of unitarity bounds for the deformation parameter $r$ and the number of allowed spacetime dimensions, $D$. For the former we show $r \geq 0$, while for the latter we derive $D \leq 26$. These results agree with the numerical evaluation of the partial-wave coefficients presented in \cite{Cheung:2023adk}. In addition to these conclusions, we derive an analytic expression for the partial-wave coefficients in general $D$ dimensions. Our representation of the coefficients is distinct, but equivalent, from the form presented in \cite{Cheung:2023adk}. Furthermore, we provide analytic expressions for the partial-wave coefficients on several Regge trajectories, and the structure of those coefficients on any Regge trajectory. These results have not appeared before.

The structure of this work is the following: in \cref{sec: generalities} we set-up our notation and conventions. We move on to \cref{sec: d4}, where we specialise the discussion in the $D=4$ case and we derive the partial-wave coefficients for the leading Regge trajectory, $a_{n,n+1}$. We, also, provide expressions for the partial-wave coefficients of the sub-leading Regge trajectories, $a_{n,n-\gamma}$, with $\gamma=\{0,1,\dots,10\}$. Finally, we re-write the original integral representation of the partial-wave coefficients as multiple sums. In \cref{sec: d} we analyse the partial-wave coefficients for general dimensions. We proceed to analyse the positivity constraints of those coefficients in \cref{sec: unitarity} and derive bounds on the parameter $r$ and the spacetime dimensions $D$. We conclude in \cref{sec: epilogue}.
\section{Generalities and setup}\label{sec: generalities}
The new infinite family of hypergeometric amplitudes is given by \cite{Cheung:2023adk}
\begin{equation}\label{eq: hyper_amplitude}
\mathcal{A}(s,t) = \frac{\Gamma(-s-1)\Gamma(-t-1)}{\Gamma(-s-t-2)}{}_3F_2\left(-s-1, -t-1, r; -s-t-2, 1+r ; 1 \right)
\,			,
\end{equation}
where in the above ${}_3F_2(a;b;z)$ is the generalised hypergeometric function\footnote{Note that relative to \cite{Cheung:2023adk} we have a shift $\{s,t\}\rightarrow\{s+1,t+1\}$. This shift precisely amounts to considering the scattering of massive states with $m^2_0=-1$, rather than massive particles.}${}^{,}$\footnote{Let us remind the reader of the definition of the generalized hypergeometric function. It is given as a formal power series; ${}_p F_{q}(\{\alpha_1,\alpha_2,\ldots,\alpha_p\};\{\beta_1, \beta_2, \ldots, \beta_q\};z)=\sum^{\infty}_{k=0}\frac{(\alpha_1)_k (\alpha_2)_k \ldots (\alpha_p)_k}{(\beta_1)_k (\beta_2)_k \ldots (\beta_q)_k} \frac{1}{k!}z^k$.}.

It is obvious that the amplitude, $\mathcal{A}(s,t)$, has poles in $s$ at $s=n=-1,0,1\ldots$ and of course the same is true for $t$.

We start by calculating the residues at the $s$ poles. We have 
\begin{equation}\label{eq: res_s}
\mathop{\mathrm{Res}}_{s = n} \mathcal{A}(s,t) = - \frac{r!}{(n+r+1)!}(t+2+r)_{n+1}
\,			.
\end{equation}
From the Gegenbauer expansion for arbitrary dimensions $D$ we know
\begin{equation}\label{eq: res_gegen}
\mathop{\mathrm{Res}}_{s = n} \mathcal{A}(s,t) = - \sum_{l=0}^{n+1} a_{n,l} C_l^{(\alpha)} \left( 1+\frac{2t}{n+4}\right)
\,			,
\end{equation}
where in the above the parameter $\alpha$ is related to the spacetime dimensions $D$ via $\alpha=\tfrac{D-3}{2}$.
Now we can just equate \cref{eq: res_s,eq: res_gegen} to get 
\begin{equation}\label{eq: res_res_gegen}
\frac{r!}{(n+r+1)!}(t+2+r)_{n+1} = \sum_{l=0}^{n+1} a_{n,l} C_l^{(\alpha)} \left( 1+\frac{2t}{n+4}\right)
\,			.
\end{equation}
We can use the fact that the Gegenbauer polynomials satisfy the following orthogonality condition
\begin{equation}\label{eq: gegen_ortho}
\int_{-1}^{+1} dx C_{\ell}^{(\alpha)} (x) C_{\ell^{\prime}}^{(\alpha)}(x)
(1-x^2)^{\alpha-\frac{1}{2}} = 2 \mathcal{K}(\ell,\alpha) \delta_{\ell \ell^{\prime}}
\,			,
\end{equation}
where in the above we have defined 
\begin{equation}
\mathcal{K}(\ell,\alpha) = \frac{\pi\Gamma(\ell+2\alpha)}{2^{2\alpha}\ell! (\ell+\alpha) \Gamma^2(\alpha)}
\,			,
\end{equation}
in order to derive the following expression for the partial-wave coefficients of \cref{eq: res_res_gegen} 
\begin{equation}\label{eq: basic_anl_relation}
	\begin{aligned}
a_{n,\ell} = 	&\frac{r!}{(n+1+r)!}\frac{1}{\mathcal{K}(\ell,\alpha)}\left[\frac{4}{(n+4)^2}\right]^{\alpha-\tfrac{1}{2}}\\
				&\int^{n+4}_0 dt C_{\ell}^{(\alpha)}\left(1-\frac{2t}{n+4}\right)(t(n+4-t))^{\alpha-\tfrac{1}{2}}(-t+2+r)_{n+1}
\,			.
	\end{aligned}
\end{equation}

By examining \cref{eq: basic_anl_relation} we make the following observations:
\begin{itemize}
\item Unlike the case of the Veneziano amplitude where $a_{n,\ell} = 0$ for $n + \ell$ equal to an even number, here we do not have that. This is due to the presence of the deformation parameter $r$. In the special case $r=0$ the coefficients are equal to $0$ as they should.
\item We have, however, that $a_{n,\ell} = 0$ for $\ell \geq n+2$, as is the case for the Veneziano amplitude as well. The vanishing of the partial-wave coefficients $a_{n,\ell}$ for the values $\ell \geq n+2$ indicates that the states that are present are of spins up to $\ell = n+1$ for a fixed mass $m_n^2$. 
\end{itemize}

We conclude this section here and discuss more the above two points in \cref{app: zero_coeffs}.
\section{The partial wave coefficients in \texorpdfstring{$D=4$}{D4} dimensions}\label{sec: d4}
Here we specialize the discussion in the $D=4$, or equivalently $\alpha=\tfrac{1}{2}$, case. \Cref{eq: basic_anl_relation} simplifies to the following expression 
\begin{equation}\label{eq: basic_anl_relation_d4}
a_{n,\ell} = 	\frac{r!}{(n+1+r)!}\frac{1+2\ell}{n+4}
				\int^{n+4}_0 dt P_{\ell}\left(1-\frac{2t}{n+4}\right)(-t+2+r)_{n+1}
\,			,
\end{equation}
where in the above $P_{\ell}(x)$ denotes the Legenedre polynomials. 

We can proceed by utilizing the generating function of the Legendre polynomials
\begin{equation}\label{eq: gen_function_legendre}
\sum^{\infty}_{\ell=0} P_{\ell}(x) t^{-\ell-1} = \frac{1}{\sqrt{1-2xt+t^2}}			\,			.
\end{equation}
and the representation of the Pochhammer symbol in terms of the Stirling number of the first kind, $s^{(b)}_a$,\footnote{in some places in the literature it is written as $s(a,b)$ but we opt for the one that is closer to the $\texttt{Mathematica}$ implementation.}  
\begin{equation}\label{eq: stirling_formula}
(x)_n = \sum^n_{k=0} (-1)^{n-k} s^{(k)}_n x^k			\,			,
\end{equation}
in order to obtain 
\begin{equation}\label{eq: def_grnkh_d4}
\sum^{\infty}_{j=0} \frac{1}{2j+1} \frac{a_{n,j}}{h^{j+1}} = \frac{1}{(n+4)(n+1)!}  \sum^{n+1}_{k=0} (-1)^{n+1-k} s^{(k)}_{n+1} \underbrace{\int^{n+4}_0 dt \frac{(-t+2+r)^k}{\sqrt{(h-1)^2 + \tfrac{4ht}{n+4}}}}_{\mathcal{G}^{(r)}_{n,k}(h)}
\,			.
\end{equation}
In the above, \cref{eq: def_grnkh_d4}, we have defined a ``pseudo generating function" $\mathcal{G}^{(r)}_{n,k}(h)$ which we can evaluate explicitly and is given by:
\begin{equation}\label{eq: general_pseudo_gen_fun_D4}
\mathcal{G}^{(r)}_{n,k}(h) = 
				\frac{1}{2^{2k+1}} \frac{(n+4)^{k+1}}{h^{k+1}} \left[(h-1)^2 + \frac{4h}{n+4}(2+r)\right]^{k}  \Big(\mathcal{E}^{+} - \mathcal{E}^{-}  \Big)				\,				,
\end{equation}
with the shorthands 
\begin{equation}
	\begin{aligned}
\mathcal{E}^{\pm}			&=		(h \pm 1) ~ {}_2 F_{1} (\tfrac{1}{2}, - k , \tfrac{3}{2} ; f^{(\pm)})
\,				,
\\
f^{(\pm)}					&= 		\frac{(h\pm1)^2}{(h-1)^2+ \tfrac{4h}{n+4}(2+r)}				
\,				.
	\end{aligned}
\end{equation}
Using the formal power-series definition for the hypergeometric function ${}_2F_{1}(a,b,c;z)$ 
\begin{equation}\label{eq: def_2F1_main}
{}_2 F_{1}(a,b,c;x) = \sum^{\infty}_{y=0} \frac{1}{y!}\frac{(a)_y(b)_y}{(c)_y} x^y			\,			,
\end{equation}
we can re-write \cref{eq: general_pseudo_gen_fun_D4} as\footnote{Note that the $\mathfrak{p}$-sum is terminated, however, this is natural since $(-a)_b=0$ holds $\forall b>a$.}
\begin{equation}\label{eq: def_grnkh_d4_conv_1}
	\begin{aligned}
\mathcal{G}^{(r)}_{n,k}(h) = 	
				&\frac{1}{2^{2k+1}} \frac{(n+4)^{k+1}}{h^{k+1}} \\
				&\sum^k_{\mathfrak{p}=0} \frac{(-k)_{\mathfrak{p}}}{(2\mathfrak{p}+1)\mathfrak{p}!} \left((h-1)^2 + \frac{4h}{n+4}(2+r)\right)^{k-\mathfrak{p}} \sum^{2\mathfrak{p}+1}_{\mathfrak{m}=0} \binom{2\mathfrak{p}+1}{\mathfrak{m}} h^{\mathfrak{m}} (1+(-1)^{\mathfrak{m}})			
\,				.
	\end{aligned}
\end{equation}
	\subsection{The leading Regge trajectory}
In \cref{eq: def_grnkh_d4_conv_1} the function $\mathcal{G}^{(r)}_{n,k}(h)$ has in total $k+1$ terms that scale according to $\tfrac{1}{h},\tfrac{1}{h^2},\dots,\tfrac{1}{h^{k+1}}$. It is a rather straightforward exercise to extract the $\tfrac{1}{h^{k+1}}$ coefficient, which is given by
\begin{equation}\label{eq: D_leading_regge_d4}
\mathcal{D}^{(r)}_{n,k} = \frac{(n+4)^{k+1}}{2^{2k+1}} \sqrt{\pi} \frac{k!}{(k+\tfrac{1}{2})!}
\,				.
\end{equation}
Let us recall at this point, that the point of the exercise is to extract the partial-wave coefficient for the leading Regge trajectory, $a_{n,n+1}$. This is the term that scales like $\tfrac{1}{h^{n+2}}$ in \cref{eq: def_grnkh_d4}. We have
\begin{equation}
\frac{1}{2n+3}a_{n,n+1} = \frac{1}{n+4} \frac{r!}{(n+r+1)!} s^{(n+1)}_{n+1} \mathcal{D}^{(r)}_{n,n+1}
\,				,
\end{equation} 
which yields 
\begin{equation}\label{eq: ann1_d4}
a_{n,n+1} = \frac{1}{4^{n+1}}\sqrt{\pi}(n+4)^{n+1}\frac{(n+1)!}{(n+\tfrac{1}{2})!}\frac{r!}{(n+r+1)!}
\,				,
\end{equation}

and \cref{eq: ann1_d4} holds for any integer $n \geq -1$. Note, also, that for $r=0$ since the amplitude reduces to the Veneziano amplitude, the partial-wave coefficient of the leading Regge trajectory, given by \cref{eq: ann1_d4}, should reduce to the result derived for that case. We have checked against the result derived in \cite{Maity:2021obe} and this is indeed the case. 
	\subsection{More on the Regge trajectories}\label{sec: regge_traj_d4}   
In this section we wish to provide some, hopefully, useful expressions for the partial-wave coefficients of the Regge trajectories. We start the discussion by considering the case $a_{n,n}$. We work in an experimental manner. That is, we compute the coefficients of interest using \cref{eq: basic_anl_relation_d4} for some low-lying values of $n$ and manage to spot the pattern. Subsequently, we perform numerous non-trivial tests of our expression. By non-trivial we mean against values for the quantum number $n$ that lie outside the data-range used to obtain the original expression. We find the following 
\begin{equation}\label{eq: ann_d4}
a_{n,n} = a_{n,n+1} ~ \frac{2n+1}{n+4} ~ 2 r 
\,					.
\end{equation}
Let us make the comment that in the $r=0$ case the coefficients given by \cref{eq: ann_d4} should coincide with those of the undeformed Veneziano amplitude. It is useful at this point to remind ourselves of the fact that for the undeformed Veneziano amplitude the partial-wave coefficients, $a_{n,\ell}$, are always equal to zero when $n+\ell$ is an even number. This agreement is manifest in \cref{eq: ann_d4}. 

Working in a similar vein for a couple more Regge trajectories, we observe that there is a striking pattern. All the coefficients can be written as 
\begin{equation}\label{eq: general_regge_d4}
a_{n,n-\gamma} = a_{n,n+1} ~ \frac{2(n-\gamma)+1}{(n+4)^{\gamma+1}} ~ \frac{1}{2} \left((-1)^{\gamma } (r-1)+r+1\right) ~ \mathcal{P}_{\gamma}(n,r) 
\end{equation}
where in the above $\mathcal{P}_{\gamma}(n,r)$ is a polynomial in $n$ and $r$. The $r$-degree of the polynomial is $\tfrac{1}{2}(-1)^{\gamma } \left((-1)^{\gamma } (2\gamma +1)-1\right)$ and only the even powers in $r$ appear. The degree in $n$ is $\tfrac{1}{4} (-1)^{\gamma } \left((-1)^{\gamma } (6\gamma +1)-1\right)$. Unfortunately, we do not have a closed-form expression for all Regge trajectories, however, we have closed-form expressions for $\gamma=\{0,1,\ldots,10\}$. We observed the general pattern using the values $\gamma=\{0,1,2,3,4\}$ and the remaining values of $\gamma$ served as checks of the pattern we found. Since the polynomials become quite unwieldy, below we provide the expressions for those that we will need in order to constrain the parameter $r$ and more expressions can be found in \cref{app: polynomials}. The form of the first 2 the 3 polynomials presented in \cref{eq: general_regge_d4_2} were obtained for values $n \leq 5$ and checked up to $n \leq 11$. The third polynomial was obtained from values $n \leq 6$ and checked against data up to $n \leq 11$.
\begin{equation}\label{eq: general_regge_d4_2}
	\begin{aligned}
\mathcal{P}_{0}(n,r)		&=		2
\,			,
\\
\mathcal{P}_{1}(n,r)		&=		
(4 n+2) r^2+\frac{n^2}{6}+\frac{19 n}{6}+\frac{23}{3}
\,			,
\\
\mathcal{P}_{2}(n,r)		&=	
\left(\frac{16n^2}{3}-\frac{4}{3}\right) r^2 + 
\frac{2n^3}{3} + \frac{43n^2}{3} + \frac{121 n}{3} + \frac{50}{3}
\,			.
	\end{aligned}
\end{equation}

Before closing the section and for concreteness and clarity we mention that the regime of validity in the formulae presented above in \cref{eq: general_regge_d4_2} for $n$ is such that the spin is never negative. More concretely for $\gamma = 0$ we have $n \geq 0$, for $\gamma = 1$ we have $n \geq 1$, and for $\gamma = 2$ we have $n \geq 2$. This logic holds for all the partial-wave coefficients.
	\subsection{The general coefficients}
It is possible, with the relations that we have derived so far, to re-write the general partial-wave coefficients given by \cref{eq: basic_anl_relation_d4}. Algorithmically we will work in the same way as above for the leading Regge trajectory, $a_{n,n+1}$. The task at hand, now, is to compute the general term that scales like $h^{-\ell-1}$ from both sides of the $h$-expansion. We re-state below what we have obtained, for the reader's convenience. The relation that we need to use in order to get the $a_{n,\ell}$ is given by:
\begin{equation}
	\begin{aligned}
\sum^{\infty}_{j=0} \frac{1}{2j+1} \frac{1}{h^{j+1}}a_{n,j} = &\frac{r!}{(n+r+1)!}\sum^{n+1}_{k=0} (-1)^{n+1-k} s^{(k)}_{n+1} \frac{1}{2^{2k+1}} (n+4)^{k}  \frac{1}{h^{k+1}}
\\
&\sum^{k}_{\mathfrak{p}=0} \frac{(-k)_{\mathfrak{p}}}{\mathfrak{p}!(2\mathfrak{p}+1)} \left[(h-1)^2 + \frac{4h}{n+4}(r+2)\right]^{k-\mathfrak{p}} \sum^{2\mathfrak{p}+1}_{\mathfrak{m}=0} \binom{2\mathfrak{p}+1}{\mathfrak{m}} h^{\mathfrak{m}} (1+(-1)^{\mathfrak{m}})		
\,			.
	\end{aligned}
\end{equation}
We will use the binomial expansion  
\begin{equation}
(a+b)^c = \sum^{c}_{d=0} \binom{c}{d}a^{c-d}b^d		
\,				,
\end{equation}
in order to re-write 
\begin{equation}
\left[(h-1)^2 + \frac{4h}{n+4}(2+r)\right]^{k-\mathfrak{p}} = 
												\sum^{k-\mathfrak{p}}_{\mathfrak{r}=0} \binom{k-\mathfrak{p}}{\mathfrak{r}} \left( \frac{4}{n+4}(2+r) \right)^{k-\mathfrak{p}-\mathfrak{r}} h^{k-\mathfrak{p}-\mathfrak{r}} (h-1)^{2\mathfrak{r}}		
\,			,
\end{equation}
and subsequently 
\begin{equation}
(h-1)^{2\mathfrak{r}} = \sum^{2\mathfrak{r}}_{\mathfrak{z}=0} \binom{2\mathfrak{r}}{\mathfrak{z}}(-1)^{2\mathfrak{r}-\mathfrak{z}}h^{\mathfrak{z}}			
\,				,
\end{equation}
thus arriving at 
\begin{equation}
	\begin{aligned}
\sum^{\infty}_{j=0} \frac{1}{2j+1} \frac{1}{h^{j+1}} a_{n,j} =					
&\frac{r!}{(n+r+1)!}\sum^{n+1}_{k=0} (-1)^{n+1-k} s^{(k)}_{n+1} \frac{1}{2^{2k+1}} (n+4)^{k} ~ \frac{1}{h^{k+1}} ~\times~ \\
&\sum^{k}_{\mathfrak{p}=0}\sum^{2\mathfrak{p}+1}_{\mathfrak{m}=0}\sum^{k-\mathfrak{p}}_{\mathfrak{r}=0} \sum^{2\mathfrak{r}}_{\mathfrak{z}=0} \binom{2\mathfrak{p}+1}{\mathfrak{m}}\binom{k-\mathfrak{p}}{\mathfrak{r}}\binom{2\mathfrak{r}}{\mathfrak{z}} \left( \frac{4}{n+4}(r+2) \right)^{k-\mathfrak{p}-\mathfrak{r}} \\
&\frac{(-k)_{\mathfrak{p}}}{\mathfrak{p}!(2\mathfrak{p}+1)} (-1)^{\mathfrak{z}} h^{\mathfrak{z}+\mathfrak{m}+k-\mathfrak{p}-\mathfrak{r}}(1+(-1)^{\mathfrak{m}})
\,				.
	\end{aligned}
\end{equation}
From the above we can read-off the terms proportional to $h^{-\ell-1}$ from both sides and the result yields\footnote{We have checked that our result, \cref{eq: general_anl_d4}, matches the expression for the partial-wave coefficients in \cite{Cheung:2023adk}, Note that there is a shift by one in the definitions of $n$ between the two works; the $n$ here has to be shifted as $n \rightarrow n+1$ to match the result in \cite{Cheung:2023adk}.}:
\begin{equation}\label{eq: general_anl_d4}
	\begin{aligned}
a_{n,\ell} = &\frac{\left( \ell+1 \right) r!}{(n+r+1)!}\sum^{n+1}_{k=0} (-1)^{n+1-k} s^{(k)}_{n+1} \frac{1}{2^{2k+1}} (n+4)^{k}  \\
								&\sum^{k}_{\mathfrak{p}=0}\sum^{2\mathfrak{p}+1}_{\mathfrak{m}=0}\sum^{k-\mathfrak{p}}_{\mathfrak{r}=0}\binom{2\mathfrak{p}+1}{\mathfrak{m}}\binom{k-\mathfrak{p}}{\mathfrak{r}}\binom{2\mathfrak{r}}{\mathfrak{p}+\mathfrak{r}-\ell-\mathfrak{m}} \frac{(-k)_{\mathfrak{p}}}{\mathfrak{p}!(2\mathfrak{p}+1)}  \\
								&\left( \frac{4}{n+4}(r+2) \right)^{k-\mathfrak{p}-\mathfrak{r}}(-1)^{\mathfrak{p}+\mathfrak{r}-\ell-\mathfrak{m}}(1+(-1)^{\mathfrak{m}})
	\end{aligned}
\end{equation}
	\subsection{A simpler expression for the general coefficients}\label{sec: appel_d4}
We will derive a simpler expression for the $a_{n,\ell}$ coefficients compared to \cref{eq: general_anl_d4}. To do so, we remind ourselves of the definition of the ``pseudo generating function" $\mathcal{G}^{(r)}_{n,k}(h)$ in $D=4$ which is given by: 
\begin{equation}\label{eq: def_gnrk_d4_appel}
\mathcal{G}^{(r)}_{n,k}(h)=\int^{n+4}_0 dt \frac{(-t+2+r)^k}{\sqrt{(h-1)^2 + \tfrac{4ht}{n+4}}}		
\,				.
\end{equation} 
As we have seen, the integral in \cref{eq: def_gnrk_d4_appel} can be performed analytically in terms of the ordinary Gauss hypergeometrics functions, ${}_2F_{1}(a,b,c;x)$. There exists another way of obtaining the integral in terms of the Appell hypergeometrics function, $F_{1}(a;b,c;d;x,y)$. The answer is given by the following:
\begin{equation}\label{eq: gen_fun_appell_d4}
\mathcal{G}^{(r)}_{n,k}(h)=2^k \frac{n+4}{h-1} F_1\left(1;-k,\tfrac{1}{2};2;\tfrac{n+4}{2+r},-\tfrac{4h}{(h-1)^2}\right)		
\,			.
\end{equation}
We recall now that the formal definition of the Appell hypergeometrics function as a power-series in the following manner
\begin{equation}\label{eq: def_F1_main}
F_{1}(a;b,c;d;x,y) = \sum^{\infty}_{e=0}\sum^{\infty}_{f=0}\frac{1}{e!}\frac{1}{f!} \frac{(a)_{e+f}(b)_{e}(c)_{f}}{(d)_{e+d}} x^{e} y^{f}			
\,			,
\end{equation}
and we also make note of the simplification: 
\begin{equation}
\frac{(1)_{x+y}}{(2)_{x+y}} = 1 + x + y
\			, 
\end{equation}
in order to obtain: 
\begin{equation}\label{eq: appell_aux02_d4}
\mathcal{G}^{(r)}_{n,k}(h)=(r+2)^k \frac{n+4}{h-1} \sum^{k}_{\mathfrak{u}=0} \sum^{\infty}_{\mathfrak{v}=0} \frac{1}{\mathfrak{u}!} \frac{1}{\mathfrak{v}!} \frac{(-k)_{\mathfrak{u}}(\tfrac{1}{2})_{\mathfrak{v}}}{1+\mathfrak{u}+\mathfrak{v}} \left(\frac{n+4}{r+2}\right)^{\mathfrak{u}} \left(-\frac{4h}{(h-1)^2}\right)^{\mathfrak{v}}				
\,				.
\end{equation}
To proceed we perform the $\mathfrak{u}$-sum in the above relation and we obtain 
\begin{equation}\label{eq: appell_aux03_d4}
\mathcal{G}^{(r)}_{n,k}(h)=(r+2)^k (n+4) \sum^{k}_{\mathfrak{v}=0} \frac{(\tfrac{1}{2})_{\mathfrak{v}}}{(\mathfrak{v}+1)!} {}_2F_{1} \left(-k,1+\mathfrak{v},2+\mathfrak{v};\tfrac{n+4}{r+2} \right) (-4)^{\mathfrak{v}} \frac{h^{\mathfrak{v}}}{(h-1)^{2\mathfrak{v}+1}}
\,			.
\end{equation}
Now, we can use, once more, the binomial expansion 
\begin{equation}\label{eq: binomial_negative}
\frac{1}{(1+x)^n} = \sum^{\infty}_{y=0} \binom{n+y-1}{y}(-1)^y x^y		
\,				,
\end{equation}
as well as the relation between the binomial coefficients and the Pochhammer symbol
\begin{equation}\label{eq: property_binom_pochammer}
\binom{x+y-1}{y} = \frac{(x)_y}{y!}			
\,				,
\end{equation} 
in order to re-write  
\begin{equation}
\frac{1}{(h-1)^{2\mathfrak{v}+1}} = (-1)^{-2\mathfrak{v} -1}\sum^{\infty}_{\mathfrak{r}=0} \frac{(2\mathfrak{v}+1)_{\mathfrak{r}}}{\mathfrak{r}!} (-1)^{2\mathfrak{r}} h^{\mathfrak{r}}
\,				,
\end{equation}
and thus the expression for $\mathcal{G}^{(r)}_{n,k}(h)$ becomes 
\begin{equation}\label{eq: appell_aux04_d4}
	\begin{aligned}
\mathcal{G}_{n,k}(h)=(r+2)^k (n+4) 	
						&\sum^{k}_{\mathfrak{v}=0} \frac{(\tfrac{1}{2})_{\mathfrak{v}}}{(\mathfrak{v}+1)!} {}_2F_{1} \left(-k,1+\mathfrak{v},2+\mathfrak{v};\tfrac{n+4}{r+2} \right) (-4)^{\mathfrak{v}} (-1)^{-2\mathfrak{v} -1} ~ h^{\mathfrak{v}} ~ \\
						&\sum^{\infty}_{\mathfrak{r}=0} \frac{(2\mathfrak{v}+1)_{\mathfrak{r}}}{\mathfrak{r}!} (-1)^{2\mathfrak{r}} h^\mathfrak{r}		
\,			.
	\end{aligned}
\end{equation}
Having obtained an explicit form as power series expansion for the ``pseudo generating function", $\mathcal{G}_{n,k}(h)$, in terms of $h$, \cref{eq: appell_aux04_d4}, it is quite straightforward to extract the coefficient $a_{n,\ell}$. It is given by\footnote{We have checked in this case as well that, \cref{eq: anl_d4_appell}, matches the expression for the partial-wave coefficients in \cite{Cheung:2023adk} as we did previously.}:
\begin{equation}\label{eq: anl_d4_appell}
	\begin{aligned}
        a_{n,\ell}=&\frac{(2\ell+1)r!}{(n+4)(n+r+1)!}
        \sum_{k=0}^{n+1}(-1)^{n+1-k} s^{(k)}_{n+1} (r+2)^k (n+4) \\
        &\sum_{\mathfrak{v}=0}^{\ell}\frac{(\frac{1}{2})_{\mathfrak{v}}}{(\mathfrak{v}+1)!} {}_2F_{1}\left(-k,1+\mathfrak{v};2+\mathfrak{v};\tfrac{n+4}{r+2}\right)	
        (-4)^{\mathfrak{v}} \frac{(2 \mathfrak{v} + 1)_{\ell-\mathfrak{v}}}{(\ell-\mathfrak{v})!}			
\,			.
	\end{aligned}
\end{equation}
Note that, while \cref{eq: anl_d4_appell}, is completely equivalent to \cref{eq: general_anl_d4} they are distinct parametrisations of the partial-wave coefficients. \Cref{eq: anl_d4_appell} comes with only two sums, instead of the four ones that appear in \cref{eq: general_anl_d4}. An argument can be made in favour of both those equations in comparison to \cref{eq: basic_anl_relation_d4} as they are just sums, rather than integration of the Legendre polynomials times non-trivial functions.  

Also, note that for $\ell=0$ we have just a single sum 
\begin{equation}\label{eq: anl_d4_appell_2}
	\begin{aligned}
        a_{n,0}=&\frac{r!}{(n+4)(n+r+1)!}
        \sum_{k=0}^{n+1}(-1)^{n+1-k} s^{(k)}_{n+1} (r+2)^k \\
        &\left(
\frac{r+2}{k+1} + (-1)^k \frac{(r+2)^{-k} (n-r+2)^{k+1}}{k+1}
         \right)		
\,			.
	\end{aligned}
\end{equation}
\section{The partial wave coefficients in \texorpdfstring{$D$}{D} dimensions}\label{sec: d}
In this section, we wish to derive similar expressions as we did previously in $D=4$ dimensions, \cref{sec: d4}, but without specifying the number of spacetime dimensions. The steps and basic relations that we need in order to manipulate the expressions have already appeared in the previous section, and hence we will proceed with a faster pace here. 

For the reader's convenience we record here again, the basic relation, as an integral over the Gegenbauers, that gives the partial-wave coefficients $a_{n,\ell}$
\begin{equation}\label{eq: pw_coefs_gen_d}
	\begin{aligned}
a_{n,\ell} = 	&\frac{r!}{(n+r+1)!} \frac{1}{\mathcal{K}(\ell,\alpha)} \frac{1}{n+4} \left(\frac{4}{(n+4)^2}\right)^{\alpha-\tfrac{1}{2}} \\
				&\int^{n+4}_0 dt C^{(\alpha)}_{\ell}\left(1 - \frac{2t}{n+4}\right) (t(n+4-t))^{\alpha-\tfrac{1}{2}}(-t+2+r)_{n+1}		
\,				.
	\end{aligned}
\end{equation}
In order to proceed, we want to make use of the generating function of the Gegenbauer polynomials
\begin{equation}\label{eq: gen_function_gegenbauer}
\sum^{\infty}_{\ell=0} C^{(\alpha)}_{\ell}(x) t^{\ell} = \frac{1}{(1-2xt+t^2)^{\alpha}}			
\,			,
\end{equation}
the representation of the Pochhammer symbol in terms of the Stirling number of the first kind, see \cref{eq: stirling_formula}, and also
\begin{equation}\label{eq: binomial_general_D_veneziano}
(n+4-t)^{\alpha-\tfrac{1}{2}} = \sum^{\infty}_{p=0} \binom{\alpha-\tfrac{1}{2}}{p}(-1)^p (n+4)^{\alpha - \tfrac{1}{2} -p}t^p		
\,			.
\end{equation}
After using the above, we obtain the following:
\begin{equation}\label{eq: G_gen_d}
	\begin{aligned}
&\sum^{\infty}_{j=0} \mathcal{K}(j,\alpha) a_{n,j} h^{j}	\\
					&= \frac{r!}{(n+r+1)!} \frac{1}{n+4} \left[ \frac{4}{(n+4)^2} \right]^{\alpha-\tfrac{1}{2}} \sum^{n+1}_{k=0} (-1)^{n+1-k} s^{(k)}_{n+1}  
					\sum^{\infty}_{p=0} \binom{\alpha-\tfrac{1}{2}}{p}(-1)^p (n+4)^{\alpha - \tfrac{1}{2} -p} \times \\
					&\underbrace{\int^{n+4}_0 dt ~ \frac{(-t+2+r)^k t^{p+\alpha-\tfrac{1}{2}}}{\left[(h-1)^2 + \frac{4ht}{n+4} \right]^{\alpha}}	}_{\mathcal{G}^{(\alpha)(r)}_{n,k,p}(h)}	\,				,
	\end{aligned}
\end{equation}
where in the above relation, \cref{eq: G_gen_d}, we have defined the ``pseudo generating function" $\mathcal{G}^{(\alpha)(r)}_{n,k,p}(h)$. The integral can be performed analytically and we obtain 
\begin{equation}\label{eq: G_appel_d}
	\begin{aligned}
\mathcal{G}^{(\alpha)(r)}_{n,k,p}(h) 
= 	
&\frac{2(r+2)^k}{2 \alpha + 2p + 1} (n+4)^{\alpha+p+\tfrac{1}{2}} \frac{1}{(h-1)^{2\alpha}} \\
&F_1 \left(\alpha+p+\tfrac{1}{2};-k,\alpha;\alpha+p+\tfrac{3}{2};\tfrac{n+4}{r+2},-\tfrac{4h}{(h-1)^2} \right)
\,				.
	\end{aligned}
\end{equation}
Now, we can proceed as we did in \cref{sec: appel_d4}, in order to re-write \cref{eq: G_appel_d} in a form that is appropriate for our manipulations. Namely, we can use the definition of the Appell hypergeometrics function as a power-series, which is given by \cref{eq: def_F1_main}, alongside with the simplification 
\begin{equation}
\frac{(\alpha+p+\tfrac{1}{2})_{x+y}}{(\alpha+p+\tfrac{3}{2})_{x+y}} = \frac{1+2\alpha+2p}{1+2\alpha+2p+2x+2y}
\,				,
\end{equation} 
and then analytically perform the $\mathfrak{u}$-sum, after which we need to use the binomial expansion, \cref{eq: binomial_negative}, and the relation between the binomial coefficients and the Pochhammer symbol given by \cref{eq: property_binom_pochammer} in order to obtain 
\begin{equation}
	\begin{aligned}
\mathcal{G}^{(\alpha)(r)}_{n,k,p}(h)
=
2(r+2)^k (n+4)^{\alpha+p+\tfrac{1}{2}} 
	&\sum^{k}_{\mathfrak{v}=0} \frac{(\alpha)_{\mathfrak{v}}}{\mathfrak{v}!} {}_2F_{1} \left(-k,\alpha+p+\tfrac{1}{2}+\mathfrak{v},\alpha+p+\tfrac{3}{2}+\mathfrak{v}; \tfrac{n+4}{r+2} \right)  \\
	&(-4)^{\mathfrak{v}}(-1)^{-2\mathfrak{v} - 2\alpha} ~ h^{\mathfrak{v}} ~ \sum^{\infty}_{\mathfrak{r}=0} \frac{(2\mathfrak{v}+2\alpha)_{\mathfrak{r}}}{\mathfrak{r}!} (-1)^{2\mathfrak{r}} h^{\mathfrak{r}}		
\,			.
	\end{aligned}
\end{equation}
After the above simplifications, the equation we need to consider in order to extract the partial-wave coefficients is given by:
\begin{equation}\label{eq: d_appell_last}
	\begin{aligned}
		\sum^{\infty}_{j=0} \mathcal{K}(j,\alpha) a_{n,j} h^{j} = &\frac{r!}{(n+r+1)!} 4^{\alpha-\tfrac{1}{2}} 
		\sum^{n+1}_{k=0} (-1)^{n+1-k} s^{(k)}_{n+1}  
		\sum^{\infty}_{p=0} \binom{\alpha-\tfrac{1}{2}}{p}(-1)^p 2 (r+)2^k \\  
		&\sum^{k}_{\mathfrak{v}=0} \frac{(\alpha)_{\mathfrak{v}}}{\mathfrak{v}!} \frac{1}{1+2\alpha+2p+2\mathfrak{v}} (-4)^{\mathfrak{v}} {}_2 F_{1} \left(-k,\alpha+p+\tfrac{1}{2}+\mathfrak{v},\alpha+p+\tfrac{3}{2}+\mathfrak{v}; \tfrac{n+4}{r+2} \right) \\
		&h^{\mathfrak{v}} \sum^{\infty}_{\mathfrak{r}=0} \frac{(2\mathfrak{v}+2\alpha)_{\mathfrak{r}}}{\mathfrak{r}!} (-1)^{2\mathfrak{r}} h^{\mathfrak{r}}		
\,			.
	\end{aligned}
\end{equation}
	\subsection{The leading Regge trajectory}
Using \cref{eq: d_appell_last} we can read-off from both sides the terms that scale as $h^{n+1}$ in order to extract the expression for the partial-wave coefficients on the leading Regge trajectory, $a_{n,n+1}$. We have that: 
\begin{equation}\label{eq: annplus1_formal}
	\begin{aligned}
a_{n,n+1} = 
&\frac{r!}{(n+r+1)!}\frac{2(r+2)^{n+1}}{\mathcal{K}(n+1,\alpha)}4^{\alpha-\tfrac{1}{2}}
\sum^{\infty}_{p=0}\sum^{n+1}_{\mathfrak{v}=0}\binom{\alpha-\tfrac{1}{2}}{p} \frac{(\alpha)_{\mathfrak{v}}}{\mathfrak{v}!}
\frac{(-1)^p(-4)^{\mathfrak{v}}}{1+2\alpha+2p+2\mathfrak{v}}   \\
&{}_2F_{1}\left(-n-1,\alpha+p+\tfrac{1}{2}+\mathfrak{v},\alpha+p+\tfrac{3}{2}+\mathfrak{v};\tfrac{n+4}{r+2}\right) \frac{(2\mathfrak{v}+2\alpha)_{n+1-\mathfrak{v}}}{(n+1-\mathfrak{v})!}		
\,			.
	\end{aligned}
\end{equation}
	\subsection{More on the Regge trajectories}
While, \cref{eq: annplus1_formal} is a formal derivation for the partial-wave coefficients of the leading Regge trajectories in arbitrary $D$-dimensions, it looks more like a simplified re-writing of the integral, rather than a helpful expression. Again, we can work experimentally in order to derive 
\begin{equation}\label{eq: annplus1_good}
a_{n,n+1} = \frac{(n+4)^{n+1}(n+1)!}{4^n} \frac{1}{(D-3)(D-1)} \frac{r!}{(n+r+1)!}\frac{1}{\left(\frac{D+1}{2}\right)_{n-1}}
\,			,
\end{equation}

for the leading Regge trajectory. Note that in deriving the above relation, \cref{eq: annplus1_good}, we used as input data the results from $4 \leq D \leq 5$ and $-1\leq n \leq 5$ and checked up to $d \leq 11$ and $n \leq 9$. It turns out that a similar behaviour to the one in the $D=4$ case is observed here as well. The sub-leading trajectories can be expressed as 
\begin{equation}\label{eq: annplus1_good_1}
a_{n,n-\gamma} = a_{n,n+1} ~ \frac{1}{(n+4)^{\gamma+1}} ~ \frac{1}{2}\left(1 + (-1)^{\gamma}(r-1)+r\right) ~ \left(D-3+2(n-\gamma)\right) ~ \mathbb{P}_{\gamma}(n,D,r)
\,			,
\end{equation}
where in the above $\mathbb{P}_{\gamma}(n,d,r)$ is a polynomial, which we do not have in a closed-form for all levels. For the first few we have the expressions: 
\begin{equation}\label{eq: annplus1_good_2}
	\begin{aligned}
\mathbb{P}_{0}(n,D,r)
&=
2
\,			,
\\
\mathbb{P}_{1}(n,D,r)
&=
\frac{12 r^2 (D+2 n-3)-D n-2 D+n^2+23 n+54}{6 (n+4)^2}
\,			,
\\
\mathbb{P}_{2}(n,D,r)
&=
\frac{(D+2 n-3) \left(-D n-2 D+n^2+25 n+58\right)+4 r^2 (D+2 n-5) (D+2 n-3)}{3 (n+4)^3}
\,			.
	\end{aligned}
\end{equation}
In order to derive the patter in $(\gamma,D)$ of \cref{eq: annplus1_good_2} we used $\gamma = \{0,1,2\}$ as input and checked against $\gamma=3$ and for the dimensions $4 \leq D \leq 7$ and checked up to $D \leq 9$. For the polynomials described in \cref{eq: annplus1_good_2} we used $n \leq 13$ and $D \leq 14$ as input and checked up to $d \leq 17$ and $D \leq 17$.
	\subsection{The general coefficients}
Finally, it is a straightforward exercise to extract the $h^{\ell}$ term from both sides of \cref{eq: d_appell_last} in order to obtain the expression for all partial-wave coefficients. It is given by\footnote{We have checked in this case also that, \cref{eq: anl_d_all}, matches the expression for the partial-wave coefficients in \cite{Cheung:2023adk} as we did in the previous cases.}:
\begin{equation}\label{eq: anl_d_all}
	\begin{aligned}
        a_{n,\ell}=
        &\frac{r!}{(n+r+1)!}\frac{1}{\mathcal{K}(\ell,\alpha)}4^{\alpha-\tfrac{1}{2}}
        \sum_{k=0}^{n+1}(-1)^{n+1-k} s^{(k)}_{n+1} (r+2)^k \sum^{\infty}_{p=0} \binom{\alpha-\tfrac{1}{2}}{p} (-1)^p \\
        &\sum_{\mathfrak{v}=0}^{\ell}\frac{(\alpha)_{\mathfrak{v}}}{\mathfrak{v}!} \frac{1}{1+2\alpha+2p+2\mathfrak{v}} {}_2F_{1}\left(-k,\alpha+p+\tfrac{1}{2}+\mathfrak{v},\alpha+p+\tfrac{3}{2}+\mathfrak{v};\tfrac{n+4}{r+2}\right)	 \\
        &(-4)^{\mathfrak{v}} \frac{(2 \mathfrak{v}+2\alpha)_{\ell-\mathfrak{v}}}{(\ell-\mathfrak{v})!}			
\,			.
	\end{aligned}
\end{equation}
\section{Comments on unitarity}\label{sec: unitarity}
The unitarity of the underlying theory, no matter what the theory, is directly related to the positivity of the partial-wave coefficients that we derived in the previous sections. The reason is that negative partial-wave coefficients indicate the exchange of ghost states. Hence, requiring that the partial-wave coefficients are non-negative numbers, is equivalent to the requirement that the theory is ghost-free.

The deformation parameter $r$ that enters \cref{eq: hyper_amplitude} can be any real number. With that in mind, we start from the expressions we derived for the $D=4$ case. From \cref{eq: ann1_d4} and for $n=0$ we obtain 
\begin{equation}
a_{0,1}=2 \frac{1}{r+1}
\,				,
\end{equation}
from which we conclude that $r>-1$ and this is our first unitarity bound. 

A second more stringent bound comes from the study of $a_{n,n}$ given by \cref{eq: ann_d4} already at $n=0$. We obtain 
\begin{equation}
a_{0,0} = r \frac{1}{r+1}
\,				,
\end{equation}
and requiring positivity of the above yields
\begin{equation}
r < -1 \lor r \geq 0
\,				.
\end{equation}
From the above, we conclude that $r\geq 0$, since the $r < -1$ solution does not allow the deformation parameter $r$ to become $0$ and thus undo the deformation of the Veneziano amplitude. The study of coefficients derived from the sub-leading trajectories, does not lead to further bounds on the allowed values of the parameter $r$. 

We continue to the general $D$ dimensions and check if we can derive any bounds on the allowed value of $D$. Here, we have already derived $r \geq 0$ and we will use this as an input. Note, also, that we are interested in integer spacetime dimensions $D$. The first non-trivial constraints in this case come from the $a_{n,n-1}$ and $n=1$. We have the expression 
\begin{equation}
a_{1,0}=\frac{12 (D-1) r^2-3 D+78}{12 (D-1) (r+1) (r+2)}
\,				,
\end{equation} 
the positivity of which leads to 
\begin{equation}
\left(r>0\land D\leq 4 (D-1) r^2+26\right)\lor 2 r\geq 1\lor D\leq 26
\,				.
\end{equation}
Clearly, the first part, which reads $\left(r>0\land D\leq 4 (D-1) r^2+26\right)$ is inconsistent as it does not allow the deformation parameter to return to the value $r=0$ and thus to obtain the un-deformed Veneziano amplitude. For the same reason we can exclude the second solution, $2 r\geq 1$, and we are only left with $D\leq 26$. Namely the underlying theory has to live below the critical dimension of string theory. Since, $r \geq 0$ one could imagine that the partial-wave coefficients of the Veneziano amplitude going to negative values above the critical dimensions of string theory could become positive for some appropriate value of $r$, however, such is not the case.

Before concluding this section we would like to add some clarifying comments. 

As we have already mentioned, in \cite{Cheung:2023adk} the authors provided an expression for all partial-wave coefficients as a double-sum. Upon explicit evaluations of the relation, they were able to derive bounds on the allowed values for the deformation parameter, $r$ and the number of spacetime dimensions, $D$. 

The results we have obtained here are in agreement with those presented in \cite{Cheung:2023adk} upon the appropriate shift that we have already mentioned in the previous section and for the special case of $m^2_0=-1$. While our formulae given by \cref{eq: general_anl_d4,eq: anl_d4_appell} for the special case of $D=4$ and \cref{eq: anl_d_all} for general-$D$ dimensions are equivalent to the result obtained in \cite{Cheung:2023adk} they are distinct parametrisations of the partial-wave coefficients. 

Furthermore, the general expressions for all partial-wave coefficients that appear both here and in \cite{Cheung:2023adk} are not expressed in terms of simple analytic functions, but rather as sums. This is more of a formal re-writing of the coefficients, rather than a straightforward expression that allows the non-negativity of the said coefficients to be manifest. Taking that into consideration, the formulae describing the partial-wave coefficients on the Regge trajectories, while not formally derived, appear to be more useful in a practical sense.
\section{Epilogue}\label{sec: epilogue}
In this work, we focused on examining the following hypergeometric deformation of the Veneziano amplitude
\begin{equation}\label{eq: final}
\mathcal{A}(s,t) = \frac{\Gamma(-s-1)\Gamma(-t-1)}{\Gamma(-s-t-2)}{}_3F_2\left(-s-1, -t-1, r; -s-t-2, 1+r ; 1 \right)
\,			,
\end{equation}
that was derived in \cite{Cheung:2023adk}. Using the decomposition into partial waves, we were able to derive bounds on the deformation parameter $r$ and the number of spacetime dimensions $D$, namely 
\begin{equation}
r \geq 0		\,		,		\qquad		\text{and}		\qquad		D \leq 26
\,				,
\end{equation} 
based on the requirement that the coefficients in the partial-wave decomposition are non-negative numbers. This requirement is a consequence of the unitarity of the underlying theory. 

We find it quite remarkable that even though the deformation parameter $r$ could be any real number at the beginning, and naively one could expect that this would not allow to derive any bounds on the dimensions of the underlying theory, the positivity of the coefficients in the expansion requires that $D$ is bounded from above. More extraordinary is the fact that this upper bound matches precisely the critical dimensions of the bosonic string. 

Unfortunately, unlike the physical intrpretation that we have on the bound of the spacetime dimensions, we do not have a physical explanation for the bound on the deformation parameter. It should, also, be noted that the allowed values of $r$ strongly depend on the value of the mass, and can be seen from the numerical analysis of \cite{Cheung:2023adk} that in the case of low masses negative values are allowed. As a concrete example, we mention that for the scattering of massless particles the parameter has to be $r \geq \tfrac{1}{2}$.

As we have mentioned already in the introduction, generalising the Veneziano amplitude is motivated from many different points of view, one of which is a question on the uniqueness of string theory. Being able to derive the critical dimension of the string from the examination of the hypergeometric Veneziano amplitude, is not a proof that string theory is unique, however, it can be seen as suggestive evidence. Of course, as was explained in \cite{Cheung:2023uwn}, one can argue that the input assumptions used as constraints to bootstrap \cref{eq: final} were neither strong nor restrictive enough, and hence it is not a big surprise that new mathematical functions were found that satisfy the bootstrap conditions.

We hope that this work is a first step towards the more systematic study of these new and exciting hypergeometric amplitudes, supplementing and extending the unitarity analysis of \cite{Cheung:2023adk}.

There are many exciting and interesting avenues for future work. 

The first and most straightforward path, would be to consider performing a similar analysis for different values of $m^2_0$, since in this work we specifically considered the case of massive scattering with $m^2_0=-1$. There is already numerical evidence from \cite{Cheung:2023adk} that for different values of $m^2_0$ the deformation parameter can be negative without violating the unitarity of the underlying theory\footnote{We are grateful to Grant Remmen for stressing this possibility to us.}. 

We still do not know the underlying theory of the amplitude given by \cref{eq: final}, if any. Taking our findings into consideration, as well as the fact that this four-point amplitude has an integral representation in terms of the Koba-Nielsen formula \cite{Cheung:2023adk}\footnote{It is worthwhile noting that from the point of view of the five-point amplitude, the Koba-Nielsen formula, the fact that a careful restriction of the parameters leads to a consistent four-point amplitude does not come as a surprise. Truly, upon an appropriate restriction of the parameters in the original Koba-Nielsen formula, in such a way that the singularities of the resulting expression are only in $s$ and $t$ and with the choice $s \leftrightarrow t$ to account for crossing symmetry, it is expected that the result is a reasonable four-point amplitude.}, perhaps the first and more natural place to try and look for an answer would be some $2 \rightarrow 2$ scattering process within string theory itself.

Furthermore, it is well-known that the Veneziano and the Virasoro-Shapiro amplitudes given by\footnote{We are considering the scattering of four tachyons of mass $\alpha^{\prime} m^2 = -1$ and $\alpha^{\prime} m^2 = -4$ in open and closed string theory respectively, and we are following conventions in which $\alpha^{\prime}=1$ for open strings and $\alpha^{\prime}=4$ for closed-string theory. We have used $s+t+u=4m^2$.}:
\begin{equation}
	\begin{aligned}
\mathcal{A}_{\text{ven}} 	&= \frac{\Gamma(-s-1)\Gamma(-t-1)}{\Gamma(-s-t-2)}
\,			,
\\
\mathcal{A}_{\text{vs}} 	&= \frac{\Gamma(-s-1)\Gamma(-t-1)\Gamma(s+t+3)}{\Gamma(s+2)\Gamma(t+2)\Gamma(-s-t-2)}
\,			,
	\end{aligned}
\end{equation}
are related via the KLT relation \cite{Kawai:1985xq}
\begin{equation}\label{eq: KLT}
\mathcal{A}_{\text{vs}} = \underbrace{\frac{\sin (\pi  s) \sin (\pi  t)}{\pi \sin (\pi  (-s-t))}}_{\text{kernel}} \mathcal{A}^2_{\text{ven}}
\,					.
\end{equation}
It would be very interesting to examine if a similar relation holds true for the hypergeometric deformations of the Veneziano and Virasoro-Shapiro amplitudes, to derive the kernel in this generalised context, and thus the generalised KLT relation.  

Finally, a straightforward path is to consider the hypergeometric Coon amplitude that was, also, derived in \cite{Cheung:2023adk} and attempt to obtain the corresponding unitarity bounds for the deformation parameters, $q$ and $r$, and perhaps the spacetime dimensions $D$ in that case. Note that this is the most general construction in terms of the hypergeometric deformations that were discussed in that article and certain limits can be taken in order to derive \cref{eq: final} from that. We believe that the bounds derived here can be used as useful input in order to derive bounds on the allowed region of values that the parameters can have in the case of the hypergeometric Coon amplitude.
\newpage
\section*{Acknowledgments} 
We are grateful to James M. Drummond for bringing \cite{Cheung:2023adk} to our attention and suggesting to carry out the unitarity analysis. 
We have greatly benefited from discussions with James M. Drummond, Pronobesh Maity, and Theodoros Nakas throughout the various stages of this project. 
We are, also, indebted to Grant Remmen, and Xinan Zhou for reading a draft of this work and offering their valuable insight and comments. 
Finally, we would like to acknowledge the hospitality of ShanghaiTech University where parts of this work were completed. 
The work of KCR is supported by starting funds from University of Chinese Academy of Sciences (UCAS), the Kavli Institute for Theoretical Sciences (KITS), and the Fundamental Research Funds for the Central Universities.
\newpage
\appendix
\section{Partial-wave coefficients that are equal to zero}\label{app: zero_coeffs}
	\subsection{The effect of a non-vanishing value for the r-parameter}
Let us discuss the first point of \cref{sec: generalities} at a bit more depth in order to showcase our argument. To do so, we remind ourselves that in the case of the Veneziano amplitude, in order to prove that $a_{n,\ell} = 0$ when $n + \ell$ is equal to an even number, we have to consider the shift $t=n+4-t^{\prime}$ \cite{Maity:2021obe}. Using the properties $(-x)_y=(-1)^y(x-y+1)_y$ and $C^{(\alpha)}_{\ell}(-x)=(-1)^{\ell}C^{(\alpha)}_{\ell}(x)$ and the integral of \cref{eq: basic_anl_relation} becomes 
\begin{equation}
	\begin{aligned}
&\int^{n+4}_0 dt C_{\ell}^{(\alpha)}\left(1-\frac{2t}{n+4}\right)(t(n+4-t))^{\alpha-\tfrac{1}{2}}(-t+2+r)_{n+1}
=
\\
-(-1)^{n+\ell}&\int^{n+4}_0 dt C_{\ell}^{(\alpha)}\left(1-\frac{2t}{n+4}\right)(t(n+4-t))^{\alpha-\tfrac{1}{2}}(-t+2-r)_{n+1}
\,				,
	\end{aligned}
\end{equation}
where in the above we renamed $t^{\prime}$ as $t$ after using the properties. Now, in the Veneziano case, which is the case $r=0$, the original integral is just re-written as $-(-1)^{n+\ell}$ times itself, and hence for $n+\ell$ any even number the result is zero. It is clear, that due to the presence of a non-zero $r$ this is no longer the case. 

Before we proceed to show that $a_{n,\ell}=0$ for $\ell \geq n+2$, we briefly mention again that this indicates the presence of states with spins up to $\ell = n+1$ at mass $m_n^2$.
	\subsection{Vanishing coefficients in \texorpdfstring{$D=4$}{D=4}}
Here we will specify the discussion to the case $D=4$. In this case, we have seen that the expression for the partial-wave coefficient simplifies drastically, see \cref{eq: basic_anl_relation_d4}. Let us focus on the integral of that expression, given by 
\begin{equation}\label{eq: app_A_01}
\int^{n+4}_0 dt P_{\ell}\left(1 - \frac{2t}{n+4} \right)(-t+2+r)_{n+1}
\,			.
\end{equation}
We will use that the Legendre polynomials satisfy 
\begin{equation}\label{eq: app_A_02}
P_{\ell}\left(1 - \frac{2t}{n+4}\right) = \sum^{\ell}_{k=0} \binom{\ell}{k}\binom{\ell+k}{k}\left(-\frac{t}{n+4}\right)^k
\,			,
\end{equation}
in order to re-write \cref{eq: app_A_01} as: 
\begin{equation}\label{eq: app_A_03}
\sum^{\ell}_{k=0} \binom{\ell}{k}\binom{\ell+k}{k}\left(-\frac{1}{n+4}\right)^k \int^{n+4}_0 dt t^k \underbrace{\left(-t+2\right)\left(-t+3\right)\ldots\left(-t+2+n+r\right)}_{(n+1)-\text{terms}}
\,			.
\end{equation}
Let us consider that $\mathfrak{a}_j$ is the coefficient of the term $t^j$ in the above and hence we have that \cref{eq: app_A_03} becomes 
\begin{equation}
\sum^{n+1}_{j=0} \mathfrak{a}_j (n+4)^{j+1} \mathcal{T}\left(\ell,j\right)
\,			,
\end{equation}
where in the above we have defined: 
\begin{equation}\label{eq: app_A_04}
\mathcal{T}(\ell,j) = \sum^{\ell}_{k=0} (-1)^k \binom{\ell}{k}\binom{\ell+k}{k} \frac{1}{k+j+1}
\,			.
\end{equation}
Notice that \cref{eq: app_A_04} can be written as: 
\begin{equation}\label{eq: app_A_05}
\mathcal{T}(\ell,j) = \int^{1}_{0} dz ~ z^j \widetilde{P}_{\ell}(z)
\,			,
\end{equation}
with $\widetilde{P}_{\ell}(z)$ being the shifted Legendre polynomial that satisfy
\begin{equation}\label{eq: app_A_06}
\widetilde{P}_{\ell}(z) = P_{\ell}(1-2z) = \sum^{\ell}_{k=0}(-1)^k \binom{\ell}{k} \binom{\ell+k}{k} z^k
\,			.
\end{equation}
Recall that the task at hand was to evaluate $\mathcal{T}(\ell,j)$. The integral in \cref{eq: app_A_05} can be evaluated to be 
\begin{equation}\label{eq: app_A_07}
\mathcal{T}(\ell,j) = \frac{(-j)_{\ell}}{(j+1)(j+2)_{\ell}}
\,			.
\end{equation}
From \cref{eq: app_A_07} we conclude that $\mathcal{T}(\ell \geq n+2,j)$ for any $j=0,1,\ldots,n+1$.
	\subsection{Vanishing coefficients in any \texorpdfstring{$D$}{D}}
Now, we proceed to compute the integral and show the vanishing of the partial-wave coefficients in any dimensions for $\ell \geq n+2$. We begin by considering the following representation of the Gegenbauer polynomials
\begin{equation}
C^{(\alpha)}_{\ell}(x) = \frac{\left(2\alpha\right)_{\ell}}{\ell!} {}_{2}F_{1}\left(-n,2\alpha+n,\alpha+\tfrac{1}{2};\tfrac{1-x}{2} \right)
\,			,
\end{equation}
which can be written in the more convenient, for our purposes, form 
\begin{equation}\label{eq: gegen_as_sum}
C^{(\alpha)}_{\ell}(x) = \frac{\left(2\alpha\right)_{\ell}}{\ell!} \sum^{\ell}_{j=0} \binom{n}{j} \frac{\left(2\alpha+n \right)_j}{\left(\alpha + \tfrac{1}{2} \right)_j} \left(\frac{x-1}{2}\right)^j
\,			.
\end{equation}
Using \cref{eq: gegen_as_sum}, the integral appearing in \cref{eq: basic_anl_relation} becomes
\begin{equation}\label{eq: zero_coeffs_1}
\int^{n+4}_0 dt \frac{\left(2\alpha\right)_{\ell}}{\ell!} \sum^{\ell}_{k=0} \binom{\ell}{k} \frac{\left(2\alpha+\ell \right)_k}{\left(\alpha + \tfrac{1}{2} \right)_k} \left(-\frac{1}{n+4}\right)^k t^k t^{\alpha-\tfrac{1}{2}}\left(-t+n+4\right)^{\alpha-\tfrac{1}{2}}\left(-t+2+r\right)_{n+1}
\,			.
\end{equation}
In the above, $\left(-t+2+r\right)_{n+1}$ has in total $(n+1)$-terms of the form:\\
$\left(-t+2+r\right)\left(-t+3+r\right)\ldots\left(-t+2+n+r\right)$. Additionally, we can use the binomial theorem to express $\left(-t+n+4\right)^{\alpha-\tfrac{1}{2}}$ as:
\begin{equation}
\left(-t+n+4\right)^{\alpha-\tfrac{1}{2}} = 
\sum^{\infty}_{p=0} \binom{\alpha-\tfrac{1}{2}}{p}(-1)^p (n+4)^{\alpha-\tfrac{1}{2}-p} t^p
\,			.
\end{equation}
Now, we consider that $a_j$ is the coefficient of $t^j$ in $\left(-t+2+r\right)\left(-t+3+r\right)\ldots\left(-t+2+n+r\right)$ and \cref{eq: zero_coeffs_1} becomes 
\begin{equation}\label{eq: zero_coeffs_2}
\frac{\left(2\alpha\right)_{\ell}}{\ell!} \sum^{n+1}_{j=0} a_j (n+4)^{j+2\alpha} \mathcal{T}(\ell,j,\alpha)
\,			,
\end{equation}
where in the above 
\begin{equation}\label{eq: zero_coeffs_3}
\mathcal{T}(\ell,j,\alpha) = 
\sum^{\ell}_{k=0}\binom{\ell}{k} \frac{\left(2\alpha+\ell \right)_k}{\left(\alpha + \tfrac{1}{2} \right)_k}(-1)^k
\sum^{\infty}_{p=0} \binom{\alpha-\tfrac{1}{2}}{p}(-1)^p \frac{2}{1+2\alpha+2j+2k+2p}
\,			.
\end{equation}
The sums in \cref{eq: zero_coeffs_3} can be performed analytically and we obtain 
\begin{equation}\label{eq: zero_coeffs_4}
\mathcal{T}(\ell,j,\alpha) = \left[ \Gamma\left(\alpha+\tfrac{1}{2}\right)\right]^2 \Gamma\left(\alpha+j+\tfrac{1}{2}\right) {}_{3}\mathcal{F}_{2}\left(\{\alpha+j+\tfrac{1}{2},-\ell,\ell+2\alpha\};\{\alpha+\tfrac{1}{2},2\alpha+j+1\};1\right)
\,			,
\end{equation}
where in the above we have used ${}_{p}\mathcal{F}_{q}(\{a_1,a_2,\ldots,a_p\};\{b_1,b_2,\ldots,b_q\};z)$ to denote the regularised hypergeometric function, which is given in terms of the generalised hypergeometric function as: 
\begin{equation}
{}_{p}\mathcal{F}_{q}(\{a_1,a_2,\ldots,a_p\};\{b_1,b_2,\ldots,b_q\};z)
=
\frac{1}{\Gamma(b_1)\Gamma(b_2)\ldots\Gamma(b_q)} {}_{p}F_{q}(\{a_1,a_2,\ldots,a_p\};\{b_1,b_2,\ldots,b_q\};z)
\,			,
\end{equation}

From \cref{eq: zero_coeffs_4} we can conclude that $\mathcal{T}(\ell \geq n+2,j,\alpha)=0$ for any number of spacetime dimensions $D$ and any $j=0,1,\ldots,n+1$.
\section{The polynomials for the Regge trajectories in \texorpdfstring{$D=4$}{D4} dimensions}\label{app: polynomials}
In this appendix we provide some additional examples for the polynomials governing the Regge trajectories in $D=4$ dimensions from \cref{sec: regge_traj_d4}. Before presenting the formulae for those, we clarify their derivations and checks. For $\mathcal{P}_{3}$ we used $n \leq 9$ and checked up to $n \leq 13$, for $\mathcal{P}_{4}$ we used $n \leq 11$ and checked up to $n \leq 13$, for $\mathcal{P}_{5}$ we used $n \leq 14$ and checked up to $n \leq 17$, for $\mathcal{P}_{6}$ we used $n \leq 16$ and checked up to $n \leq 18$, for $\mathcal{P}_{7}$ we used $n \leq 19$ and checked up to $n \leq 22$, for $\mathcal{P}_{8}$ we used $n \leq 21$ and checked up to $n \leq 25$, for $\mathcal{P}_{9}$ we used $n \leq 22$ and checked up to $n \leq 24$, and for $\mathcal{P}_{10}$ we used $n \leq 25$ and checked up to $n \leq 28$.

\begin{equation}\label{eq: general_regge_d4_app}
	\begin{alignedat}{3}
\mathcal{P}_{3}(n,r)		=		
&\left(\frac{16 n^3}{3}-8 n^2-\frac{4 n}{3}+2\right) r^4
+
\left(\frac{4 n^4}{3}+\frac{92n^3}{3}+\frac{215 n^2}{3}-\frac{23 n}{3}-18\right) r^2
+
\\
&\frac{n^5}{36}+\frac{401 n^4}{360}+\frac{1367 n^3}{90}+\frac{23503 n^2}{360}+\frac{5923 n}{60}+\frac{523}{15}
\,			,
\\
\mathcal{P}_{4}(n,r)		=
&\left(\frac{64 n^4}{15}-\frac{256 n^3}{15}+\frac{224 n^2}{15}+\frac{64n}{15}-4\right) r^4
+
\\
&\left(\frac{16 n^5}{9}+\frac{376 n^4}{9}+36 n^3-\frac{1486 n^2}{9}-\frac{82 n}{9}+\frac{116}{3}\right) r^2
+
\\
&\frac{n^6}{9}+\frac{218 n^5}{45}+\frac{4157 n^4}{60}+\frac{24743 n^3}{90}+\frac{55921 n^2}{180}-\frac{2071 n}{30}-82
\,			,
\\
\mathcal{P}_{5}(n,r)		=	
&\left(\frac{128 n^5}{45}-\frac{64 n^4}{3}+\frac{448 n^3}{9}-32 n^2-\frac{568 n}{45}+\frac{28}{3}\right) r^6 
+
\\
&\left(\frac{16 n^6}{9}+\frac{368 n^5}{9}-\frac{680 n^4}{9}-\frac{2440 n^3}{9}+\frac{3889 n^2}{9}+\frac{587 n}{9}-\frac{310}{3}\right) r^4
+
\\
&\left(\frac{2 n^7}{9}+\frac{461 n^6}{45}+\frac{13139 n^5}{90}+\frac{14515 n^4}{36}-\frac{2221 n^3}{9}-\frac{224179 n^2}{180}+\frac{1577 n}{30}+286\right) r^2
+
\\
&\frac{n^8}{324}+\frac{103 n^7}{540}+\frac{72011 n^6}{15120}+\frac{31553 n^5}{560}+\frac{285899 n^4}{1008}+\frac{1033789 n^3}{1680}+
\\
&\frac{4928053n^2}{11340}-\frac{594829 n}{3780}-\frac{1138}{9}
\,			,
\\
\mathcal{P}_{6}(n,r)		=
&\left(\frac{512 n^6}{315}-\frac{2048 n^5}{105}+\frac{5248 n^4}{63}-\frac{1024 n^3}{7}+\frac{23648 n^2}{315}+\frac{3968 n}{105}-24\right) r^6	
+
\\
&\left(\frac{64 n^7}{45}+\frac{1376 n^6}{45}-\frac{8576 n^5}{45}+\frac{16 n^4}{9}+\frac{52756 n^3}{45}-\frac{55546 n^2}{45}-\frac{1406 n}{5}+308\right) r^4
+
\\
&\left(\frac{8 n^8}{27}+\frac{1904 n^7}{135}+\frac{5108 n^6}{27}+\frac{8144 n^5}{135}-\frac{105671 n^4}{54}-\frac{126079 n^3}{135}+ \right.
\\
&\left.
\frac{87601 n^2}{18}+\frac{10327n}{45}-\frac{3292}{3}\right) r^2
+
\\
&\frac{n^9}{81}+\frac{221 n^8}{270}+\frac{16141 n^7}{756}+\frac{69851 n^6}{280}+\frac{2588081 n^5}{2520}+
\\
&\frac{272901 n^4}{280}-\frac{11443967 n^3}{4536}-\frac{7777853n^2}{1890}+\frac{356711 n}{630}+964
\,			,
	\end{alignedat}
\end{equation}

\begin{equation}\label{eq: general_regge_d4_app_2}
	\begin{alignedat}{3}
\mathcal{P}_{7}(n,r)		=
&\left(\frac{256 n^7}{315}-\frac{128 n^6}{9}+\frac{4288 n^5}{45}-\frac{2720 n^4}{9}+\frac{19792 n^3}{45}-\frac{1688 n^2}{9}-\frac{12172 n}{105}+66\right) r^8
+
\\
&\left(\frac{128 n^8}{135}+\frac{2432 n^7}{135}-\frac{32096 n^6}{135}+\frac{16864 n^5}{27}+
\right.
\\
&\left.
\frac{107992 n^4}{135}-\frac{620152 n^3}{135}+\frac{504206
   n^2}{135}+\frac{9982 n}{9}-980\right) r^6
+
\\
&\left(\frac{8 n^9}{27}+\frac{1924 n^8}{135}+\frac{22508 n^7}{135}-\frac{94526 n^6}{135}-\frac{677699 n^5}{270}+
\right.
\\
&\left.
\frac{3716929 n^4}{540}+\frac{1240327
   n^3}{135}-\frac{3456947 n^2}{180}-\frac{64279 n}{30}+4382\right) r^4	
+
\\
&\left(\frac{2 n^{10}}{81}+\frac{698 n^9}{405}+\frac{8641 n^8}{189}+\frac{468323 n^7}{945}+
\right.
\\
&\left.
\frac{75233 n^6}{72}-\frac{1451983 n^5}{360}-\frac{8123005
   n^4}{648}+\frac{18345539 n^3}{3240}+
\right.
\\
&\left.   
\frac{10273985 n^2}{378}-\frac{737869 n}{630}-6028\right) r^2
+ 
\\
&\frac{n^{11}}{3888}+\frac{841 n^{10}}{38880}+\frac{2152081 n^9}{2721600}+\frac{5665843 n^8}{362880}+\frac{5406743 n^7}{32400}+\frac{30769397n^6}{36288}
+
\\
&\frac{972735997 n^5}{544320}-\frac{38975383 n^4}{1088640}-\frac{7255356281 n^3}{1360800}-\frac{460545103 n^2}{90720}+\frac{15354599n}{12600}+\frac{18871}{15}
\,			,
	\end{alignedat}
\end{equation}

\begin{equation}\label{eq: general_regge_d4_app_3}
	\begin{alignedat}{3}
\mathcal{P}_{8}(n,r)		=
&\left(\frac{1024 n^8}{2835}-\frac{8192 n^7}{945}+\frac{11264 n^6}{135}-\frac{2048 n^5}{5}+\frac{144256 n^4}{135}-\frac{60928 n^3}{45}
+
\right.
\\
&\left.
\frac{1390016n^2}{2835}+\frac{344192 n}{945}-\frac{572}{3}\right) r^8	
+
\\
&\left(\frac{512 n^9}{945}+\frac{7936 n^8}{945}-\frac{13184 n^7}{63}+\frac{53696 n^6}{45}-\frac{74464 n^5}{45}-\frac{241136 n^4}{45}+
\right.
\\
&\left.
\frac{3306440n^3}{189}-\frac{11112316 n^2}{945}-\frac{1344076 n}{315}+3256\right) r^6 
+
\\
&\left(\frac{32 n^{10}}{135}+\frac{7616 n^9}{675}+\frac{7592 n^8}{75}-\frac{303824 n^7}{225}+\frac{134866 n^6}{225}+
\right.
\\
&\left.
\frac{3756572 n^5}{225}-\frac{27364001n^4}{1350}-\frac{36813233 n^3}{675}+\frac{34407883 n^2}{450}+\frac{945907 n}{75}-\frac{89292}{5}\right) r^4
+
\\
&\left(\frac{8 n^{11}}{243}+\frac{964 n^{10}}{405}+\frac{531536 n^9}{8505}+\frac{1619614 n^8}{2835}-\frac{1250071 n^7}{1134}-\frac{2318551 n^6}{180}+
\right.
\\
&\left.
\frac{12720941
   n^5}{4860}+\frac{133287227 n^4}{1620}+\frac{77483591 n^3}{6804}-\frac{449387851 n^2}{2835}-\frac{2828951 n}{945}+\frac{104056}{3}\right) r^2
+
\\
&\frac{n^{12}}{972}+\frac{112 n^{11}}{1215}+\frac{1196773 n^{10}}{340200}+\frac{11905489 n^9}{170100}+\frac{70047469 n^8}{100800}+
\\
&
\frac{548246269n^7}{226800}-\frac{66696559 n^6}{38880}-\frac{48406423 n^5}{1944}-\frac{79199036753 n^4}{2721600}+\frac{31022789003 n^3}{680400}+
\\
&\frac{17353447943n^2}{226800}-\frac{20749049 n}{2100}-\frac{259286}{15}
\,			,	
	\end{alignedat}
\end{equation}

\begin{equation}\label{eq: general_regge_d4_app_4}
	\begin{alignedat}{3}
\mathcal{P}_{9}(n,r)		=
&\left(\frac{2048 n^9}{14175}-\frac{1024 n^8}{225}+\frac{280576 n^7}{4725}-\frac{93184 n^6}{225}+
\right.
\\
&\left.
\frac{1117952 n^5}{675}-\frac{843136 n^4}{225}+\frac{60356992n^3}{14175}-\frac{99392 n^2}{75}-\frac{368216 n}{315}+572\right) r^{10}
+
\\
&\left(\frac{256 n^{10}}{945}+\frac{2816 n^9}{945}-\frac{15104 n^8}{105}+\frac{433408 n^7}{315}-\frac{228128 n^6}{45}+\frac{138272 n^5}{45}+
\right.
\\
&\left.
\frac{25833776n^4}{945}-\frac{62429264 n^3}{945}+\frac{12001739 n^2}{315}+\frac{1711723 n}{105}-11154\right) r^8
+
\\
&\left(\frac{64 n^{11}}{405}+\frac{14752 n^{10}}{2025}+\frac{24752 n^9}{675}-\frac{1002248 n^8}{675}+\frac{4159412 n^7}{675}+
\right.
\\
&\left.
\frac{6835234 n^6}{675}-\frac{172516181n^5}{2025}+\frac{181332943 n^4}{4050}+
\right.
\\
&\left.
\frac{189778778 n^3}{675}-\frac{413439001 n^2}{1350}-\frac{2927713 n}{45}+73612\right) r^6
+
\\
&\left(\frac{8 n^{12}}{243}+\frac{328 n^{11}}{135}+\frac{521078 n^{10}}{8505}+\frac{1100114 n^9}{2835}-\frac{25659041 n^8}{5670}-\frac{19007911n^7}{1890}+
\right.
\\
&\left.
\frac{145457699 n^6}{1944}+\frac{93677327 n^5}{1080}-\frac{30023039233 n^4}{68040}-\frac{5731391107 n^3}{22680}+
\right.
\\
&\left.
\frac{1636399621n^2}{1890}+\frac{36484051 n}{630}-190028\right) r^4
+
\\
&\left(\frac{n^{13}}{486}+\frac{941 n^{12}}{4860}+\frac{1288333 n^{11}}{170100}+\frac{50049121 n^{10}}{340200}+\frac{557418061 n^9}{453600}+
\right.
\\
&\left.
\frac{137847077n^8}{907200}-\frac{933350689 n^7}{34020}-\frac{2027689027 n^6}{38880}+\frac{212491115287 n^5}{1360800}+
\right.
\\
&\left.
\frac{1087556641399 n^4}{2721600}-\frac{50810550289n^3}{226800}-\frac{2094388529 n^2}{2800}+\frac{11761469 n}{252}+162838\right) r^2
+
\\
&\frac{n^{14}}{58320}+\frac{107 n^{13}}{58320}+\frac{362693 n^{12}}{4082400}+\frac{10121317 n^{11}}{4082400}+\frac{1667002441 n^{10}}{39916800}+\frac{3207436571n^9}{7983360}+
\\
&\frac{163569574211 n^8}{89812800}+\frac{68609574763 n^7}{35925120}-\frac{4576464369611 n^6}{359251200}-\frac{14953602874033n^5}{359251200}-
\\
&\frac{927033772163 n^4}{59875200}+\frac{2596960946917 n^3}{29937600}+\frac{441409885271 n^2}{4989600}-\frac{2649524531 n}{138600}-\frac{314354}{15}
\,			,
	\end{alignedat}
\end{equation}

\begin{equation}\label{eq: general_regge_d4_app_5}
	\begin{alignedat}{3}
\mathcal{P}_{10}(n,r)		=
&\left(\frac{8192 n^{10}}{155925}-\frac{65536 n^9}{31185}+\frac{370688 n^8}{10395}-\frac{3473408 n^7}{10395}+\frac{13976576 n^6}{7425}-\frac{1925120n^5}{297}
+
\right.
\\
&\left.
\frac{409485056 n^4}{31185}-\frac{425455616 n^3}{31185}+\frac{63601568 n^2}{17325}+\frac{13240064 n}{3465}-1768\right) r^{10}
+
\\
&\left(\frac{1024 n^{11}}{8505}+\frac{5632 n^{10}}{8505}-\frac{137728 n^9}{1701}+\frac{666112 n^8}{567}-\frac{2309504 n^7}{315}+\frac{7955776 n^6}{405}
+
\right.
\\
&\left.
\frac{2012032n^5}{1701}-\frac{214121696 n^4}{1701}+\frac{2118827684 n^3}{8505}-\frac{119186042 n^2}{945}-\frac{11763526 n}{189}+\frac{117260}{3}\right) r^8
+
\\
&\left(\frac{256 n^{12}}{2835}+\frac{55808 n^{11}}{14175}-\frac{24064 n^{10}}{14175}-\frac{1096192 n^9}{945}+\frac{1090592 n^8}{105}-\frac{92613952n^7}{4725}-
\right.
\\
&\left.
\frac{1380237872 n^6}{14175}+\frac{1098647552 n^5}{2835}-\frac{8138435 n^4}{567}-\frac{2140478078 n^3}{1575}+\frac{646434493 n^2}{525}+
\right.
\\
&\left.
\frac{33164102n}{105}-305448\right) r^6
+
\\
&\left(\frac{32 n^{13}}{1215}+\frac{11888 n^{12}}{6075}+\frac{1942184 n^{11}}{42525}+\frac{3789124 n^{10}}{42525}-\frac{6254986 n^9}{945}+\frac{200483701n^8}{14175}+
\right.
\\
&\left.
\frac{10897731679 n^7}{85050}-\frac{7691537891 n^6}{24300}-\frac{30358969699 n^5}{34020}+\frac{359628196349 n^4}{170100}+
\right.
\\
&\left.
\frac{122628506327n^3}{56700}-\frac{21427008719 n^2}{4725}-\frac{153371047 n}{315}+1006456\right) r^4
+
\\
&\left(\frac{2 n^{14}}{729}+\frac{976 n^{13}}{3645}+\frac{897097 n^{12}}{85050}+\frac{981730 n^{11}}{5103}+\frac{1102788149 n^{10}}{1020600}-\frac{15745694n^9}{1701}-
\right.
\\
&\left.
\frac{207637186087 n^8}{4082400}+\frac{25174204033 n^7}{204120}+\frac{157940397331 n^6}{226800}-\frac{10725272695 n^5}{20412}-
\right.
\\
&\left.
\frac{14608972386719n^4}{4082400}+\frac{10269635113 n^3}{22680}+\frac{681990515411 n^2}{113400}-\frac{17273033 n}{210}-1290532\right) r^2
+
\\
&\frac{n^{15}}{14580}+\frac{227 n^{14}}{29160}+\frac{11153 n^{13}}{28350}+\frac{22796371 n^{12}}{2041200}+\frac{16378424261 n^{11}}{89812800}+\frac{89710139017n^{10}}{59875200}+
\\
&\frac{488830522129 n^9}{179625600}-\frac{1146948964289 n^8}{44906400}-\frac{1733919836167 n^7}{14968800}+\frac{1953314586743n^6}{179625600}+
\\
&\frac{137895712720613 n^5}{179625600}+\frac{8473021434481 n^4}{9979200}-\frac{2101852716961 n^3}{1663200}-\frac{1626950129029n^2}{831600}+
\\
&\frac{747757867 n}{2772}+435964
\,			.
	\end{alignedat}
\end{equation}
\newpage
\bibliographystyle{ssg}
\bibliography{hyperamplitude}
\end{document}